\documentclass[onecolumn]{webofc}

\usepackage[varg]{txfonts}   
\def\m{\mu}
\def\n{\nu}

\def\){\right)}
\def\({\left( }
\def\]{\right] }
\def\[{\left[ }

\newcommand{\be}{\begin{equation}}
\newcommand{\ee}{\end{equation}}

\def\bsk{{\boldsymbol{k}}}

\def\bsq{{\boldsymbol{q}}}

\def\a{\alpha}

\begin{document}
\title{Wee partons in QCD and gravity: double copy and universality}
%
%

\author{\firstname{Himanshu} \lastname{Raj}\inst{1}\fnsep\thanks{\email{himanshu.raj@stonybrook.edu}} \and
        \firstname{Raju} \lastname{Venugopalan}\inst{2,1}\fnsep\thanks{\email{rajuv@bnl.gov}} 
}

\institute{Center for Frontiers in Nuclear Science, Department of Physics and Astronomy, Stony Brook University, NY 11794, USA
\and
           Physics Department, Brookhaven National Laboratory, Upton, NY 11973, USA 
          }

\abstract
  {We discuss a quantitative "double copy" between radiation from shockwave collisions in Einstein gravity and in QCD.  The correspondence extends to $2\rightarrow N$ amplitudes in Regge asymptotics. The classicalization and unitarization of these amplitudes at maximal occupancy, corresponding to  black hole and Color Glass Condensate (CGC) states respectively, are described by the emergent Goldstone dynamics of wee partons. We outline some  consequences of the universal dynamics on both sides of the correspondence. 
   }
\maketitle
%
A quantitative color-kinematics duality between amplitudes in perturbative QCD (pQCD) and amplitudes in Einstein gravity was discovered by Bern, Carrasco and Johansson (BCJ) \cite{Bern}. Though not as widely known, a double copy relation between gravitational amplitudes and QCD amplitudes in high energy Regge asymptotics of both theories was derived previousy by  Lipatov~\cite{Lipatov}, showing that the effective gravitational vertex for the $2\rightarrow 3$ scattering amplitude is (for exchanged $t$-channel transverse momenta $\bsq_{1,2}$  whose four-momenta satisfy $q_1+q_2=k$, with $k$ the four-momentum of the radiated graviton)
\begin{equation}
\label{LipatovVertexGrav}
    \Gamma^{\m\n}(\bsq_1, \bsq_2) = \frac12 \left(C^\m C^\n - N^\m N^\n \right)~,
\end{equation}
with $C^\mu$ the QCD Lipatov vertex and $N^\mu$  the QED soft photon bremsstrahlung vertex. They are key elements of 2-D effective theories describing the Regge asymptotics of $2\rightarrow N$ amplitudes in both QCD and gravity ~\cite{Amati}. Explicit expressions for the vertices are given in \cite{Raj:2023iqn}.

The QCD Lipatov vertex is obtained from solutions of Yang-Mills (YM) equations~\cite{Blaizot} in the Color Glass Condensate effective field theory (CGC EFT)~\cite{Gelis:2010nm}, where hadron-hadron collisions are replaced by collisions of gluon shockwaves. The radiation field is given by 
\begin{align}
\label{QCDLVLCEik1}
    A^{\mu,a}(k) = -\frac{g^3}{k^2}\int  \frac{d^2\bsq_{2}}{\(2\pi\)^2} \frac{{\tilde\rho}_A(\bsq_1)}{\bsq_{1}^2} \frac{{\tilde\rho}_B(\bsq_2)}{\bsq_{2}^2} T^b T^c i f^{a b c} \,C^\m(\bsq_1, \bsq_2)~.
\end{align}
Here ${\tilde \rho}_A$, ${\tilde \rho}_B$ 
are Fourier transforms of the number density of the colliding classical color charges. Further, we employ a "dilute-dilute" approximation where typical momentum transfers are much larger than the respective saturation scales (to be discussed shortly). 

In \cite{Raj:2023irr}, we showed that an analogous computation of the inclusive gravitational wave field produced in the collision of shockwaves in Einstein gravity can be expressed as 
\begin{equation}
\label{GravityLVLCEik1}
    A^{\mu\nu}(k) = \frac{\kappa^3 s}{2 k^2}\int  \frac{d^2\bsq_{2}}{\(2\pi\)^2} \frac{{\tilde \rho}_A(\bsq_1)}{\bsq_{1}^2} \frac{{\tilde\rho}_B(\bsq_2)}{\bsq_{2}^2} \, \Gamma^{\m\n}(\bsq_1, \bsq_2) ~,
\end{equation}
where $\Gamma^{\mu\nu}$ is Lipatov's result in Eq.~\eqref{LipatovVertexGrav}, $\kappa$ is proportional to Newton's constant and $s$ is the squared center-of-mass energy. This confirms in the shockwave picture the remarkable correspondence uncovered by Lipatov. 

Recovering the  Lipatov duality from the BCJ double copy requires extensions of the gauge theory~\cite{SabioVera}.  Instead, a classical double copy between perturbative solutions of YM equations and Einstein's equations~\cite{Monteiro:2014cda,Goldberger:2016iau} maps  the leading order radiation field from classical color charged particles in a slowly varying gauge field background~\cite{Wong:1970fu}  to gravitational radiation  with the following color-kinematic replacements \cite{Goldberger:2016iau}: $c^a_\a \to p^\m_\a$, $g \to \kappa$, 
$$i f^{a_1 a_2 a_3} \rightarrow \Gamma^{\nu_1 \nu_2 \nu_3}\left(q_1, q_2, q_3\right)=
    -\frac{1}{2}\left(\eta^{\nu_1 \nu_3}\left(q_1-q_3\right)^{\nu_2}+\eta^{\nu_1 \nu_2}\left(q_2-q_1\right)^{\nu_3}+\eta^{\nu_2 \nu_3}\left(q_3q_2\right)^{\nu_1}\right).$$
Here $c^a$ represents the color charge of massive particles with arbitrary four-velocity $v_\mu$. Taking the ultrarelativistic limit of the YM radiation field in \cite{Goldberger:2016iau} recovers Eq.~\eqref{QCDLVLCEik1}. Further, making the above classical double copy replacement  in the general expression given in \cite{Goldberger:2016iau}  and appropriately taking the ultrarelativistic limit~\cite{Raj:2023iqn}, one recovers Eq.~\eqref{GravityLVLCEik1}. 

The above results suggest that strong field techniques developed for the Regge regime of high occupancies may have a quantitative correspondence to similar dynamics in gravity. In pQCD, $2\rightarrow N+2$ scattering is described by the BFKL equation~\cite{BFKL}, with rapidities of the $2+N$ hard particles ordered as  $y_0^+ \gg y_1^+ \gg y_2^+ \gg \cdots \gg y_N^+ \gg y_{N+1}^+$ and transverse momenta  $|\bsk_i| \simeq |\bsk|\gg \Lambda_{\rm QCD}$. The cross-section is dominated by the slowest "wee" gluons with $x\ll 1$ ($y= Ln(x)$); its multiplicity grows as $\simeq e^{\lambda y}$, where the growth rate $\lambda$ can be computed explicitly. In gravity, as estimated by Lipatov, the growth is far more rapid~\cite{Lipatov}. 

In QCD,  the rapid growth in wee gluon distributions leads to gluon saturation, a nonperturbative state of maximal occupancy $\sim 1/\alpha_s(Q_S)$ characterized by a semi-hard saturation scale $Q_S(x)$. 
For recent developments relevant to these proceedings, see \cite{Mantysaari:2023gkw}. 
The Black Hole N-Portrait (BHNP)~\cite{Dvali:2011aa} similarly describes black holes as self-bound overoccupied states of wee gravitons. Here the likelihood of self-bound classical lumps in $2\rightarrow N$ scattering is treated from an information theory perspective. The probability of creating a high occupancy state of gravitons with $N = 1/\alpha$ is $P_{2\rightarrow N} \sim e^S\, e^{-1/\alpha}$, where $\alpha(Q) = Q^2/M_{Planck}^2$ is the interaction strength for momentum transfer $Q$.  Occupancies $N=1/\alpha$ thus correspond to the gravitation  saturation scale $Q\equiv Q_S$.  Clearly  $P_{2\rightarrow N}=O(1)$ only when the entropy (the logarithm of the number of nearly degenerate microstates) satisfies $S=1/\alpha$. 
 
 In \cite{Dvali:2019jjw}, it was conjectured that all such high occupancy states satisfy the area law 
 \begin{equation}
 S =  \frac{1}{\alpha} =N = {\rm Area} \times f_G^2\,,
 \label{entropy}
 \end{equation}
 where the area is that of the classical lump with occupancy $N=1/\alpha$ created in the $2\rightarrow N$ process. For a 4-D black hole where 
 $N= M_{BH}^2/M_{Planck}^2$ (with $M_{BH}$ and $M_{Planck}$ respectively the black hole mass and the Planck mass),  $f_G= M_{Planck} = 1/\sqrt{G}$ is the Goldstone scale representing the breaking of scale invariance. Defining the Schwarzchild radius $R_S= 1/Q_S$  recovers parametrically the well known relation $R_S = 2\,G\,M_{BH}$. Thus the rightmost equality of Eq.~\eqref{entropy} is equivalent to the leftmost equality when $N=1/\alpha$.  The leftmost equality saturates Bekenstein's bound for the entropy $S \leq 2\pi E R$ representing information  packed in a region of radius $R$ for a given energy $E$. In our case, $E = N/R_S$, where $R_S$ saturates the bound. The rightmost equality saturates the Bekenstein-Hawking bound for a black hole. 
 
 Thus classicalization and perturbative unitarization (henceforth referred to as criticality) of  $2\rightarrow N$ graviton amplitudes
 occurs for $\alpha\, N=1$, saturating the Bekenstein entropy bound.  The dynamics of CGC states and that of black holes are indistinguishable at the critical boundary $\alpha \,N=1$~\cite{Dvali:2021ooc}.  Just as black holes screens information at distance scales less than $R_S$, the QCD saturation scale $Q_S$ screens color for distances $> 1/Q_S$.  Away from criticality, the dynamics of the two theories are completely different: QCD is strongly coupled in the infrared and weakly coupled in the ultraviolet- the reverse is true in gravity. 
 
 In the CGC EFT, the classical lump  is a "shockwave" localized on the light cone comprised of the large occupancy of wee partons in a hadron distributed around faster partons.  It breaks translational invariance and a global sub-group of color since either side of the shockwave corresponds to differing pure gauges~\cite{MV}. The resulting Goldstone field is the highly occupied field $A^{\mu,a}$ of a single shockwave which, in the CGC framework, is static in light cone time;
 however in general, a time scale must control their decay~\cite{Coleman:1977hd}. Kinematic considerations  dictate $\tau_K \sim P^+/Q^2$, where $P^+$ is the large hadron light cone momentum. The lump's semi-classical nature, with $\Delta \varepsilon \sim Q_S/N$, where $\Delta \varepsilon$ is the spacing of wee parton energy levels, suggests instead the decay time  $\tau_E = N/Q_S\equiv  1/(\alpha_S\,Q_S)$~\cite{Dvali:2019jjw,Dvali:2021ooc}. Hence $\tau_E \ll \tau_K$. 
 
 The scale governing  the Goldstone decay constant  of the classical lump in QCD is 
 $f_G^2 = (R_S\partial_- A^a) (R_S\partial_+ A^a)\sim N/R_S^2$, since the magnitude of the field is $O(\sqrt{N}/R_S)$ and the variation in the lifetime $\partial_- = \partial_+ \sim 1/R_S$; equivalently, wee Goldstone modes have  $q^+_1=q^-_2 \sim Q_S$.  In pQCD language, we include dynamics beyond Glauber gluon modes. The lifetime of the classical lump in terms of the Goldstone scattering rate is  $\tau_G= (\Gamma_G)^{-1} = (\sigma\,n\,N)^{-1} = (\alpha_S Q_S)^{-1}$, which recovers $\tau_G=\tau_E$. Here the cross-section $\sigma = \alpha_S^2/f_G^2$, the density $n= N\,Q_S^3$, and $N=1/\alpha_S$, the phase space occupancy of wee partons at criticality. Interestingly, this scale $f_G$ is also obtained from the entanglement entropy of soft gluons  \cite{Duan:2021clk}.  
 
The Goldstone dynamics of black holes and the CGC is argued to be universal at criticality \cite{Dvali:2021ooc}, suggested by identical three-gluon and three-graviton couplings. For the former, $g\, f^{abc}\partial \sim \sqrt{\alpha_S} Q_S = Q_S^2/f_G$. For the latter, the classical double copy gives $\kappa\,\partial\cdot \partial\rightarrow \partial^2/M_{Planck} = Q_S^2/f_G$, recalling that $f_G= M_{Planck}$ in gravity. Not least, the Goldstone picture~\cite{Dvali:2021ooc} of the decay of the semi-classical CGC lump is implicit in the bottom-up thermalization scenario of the quark-gluon plasma~\cite{Bottomup}. The Goldstone decay of the CGC shockwave generates final state effects modifying the emission of soft gluons and photons in DIS and in p+A collisions. 
Another interesting possibility is a Goldstone kinetic theory of a novel infrared cascade in heavy-ion collisions providing a microscopic description of 
stringy large scale structure seen in numerical simulations of 3+1-D YM equations~\cite{String-Tension}.

We end by discussing implications of the classical double copy and the CGC EFT for gravitational wave distributions and black hole formation. 
In the CGC,  the shockwave gluon field and  quark and gluon propagators in this  background~\cite{Shock} can be employed to derive a powerful nonlinear renormalization group (RG)~\cite{JIMWLK} description of $2\rightarrow N$ scattering, generalizing BFKL to the saturation regime. In particular, criticality as we have defined it, can be understood as a nontrivial fixed point of the simplest realization of this framework, the Balitsky-Kovchegov equation~\cite{BK}. Can one understand black hole states as fixed points of an analogous RG equation at a critical impact parameter $b_c=R_S$? Can we extract from data at future gravitational wave observatories, variations in the gravitational wave spectrum (computed as a function of impact parameter and frequency), in close black hole encounters~\cite{Escriva:2022duf}? These, and related questions, will be addressed in forthcoming work. 

\section*{Acknowledgements}
We thank Juergen Berges, Gia Dvali, Aleksas Mazeliauskas, Thimo Preis, Jan Pawlowski and Parmeswaran Nair for useful discussions. R.V is supported by the U.S. Department of Energy, Office of Science under contract DE-SC0012704  within the framework of the SURGE Topical Theory Collaboration, and by an LDRD at BNL. He is also supported by the Simons Foundation under Award number 994318 (Simons Collaboration on Confinement and QCD Strings). R.V  thanks the DFG  SFB 1225 (ISOQUANT) at Heidelberg University for hospitality and support. H.R is supported by Simons Foundation Award 994318.


\begin{thebibliography}{}
\bibitem{Bern}
Z.~Bern, J.J.M. Carrasco, H.~Johansson, Phys. Rev. D \textbf{78}, 085011
  (2008).
 \bibitem{Lipatov}
L.N. Lipatov, Sov. Phys. JETP \textbf{55}, 582 (1982); L.N. Lipatov, Phys. Lett. B \textbf{116}, 411 (1982)
\bibitem{Amati}
D.~Amati, M.~Ciafaloni, G.~Veneziano, Int. J. Mod. Phys. A \textbf{3}, 1615
  (1988); L.N. Lipatov, Nucl. Phys. B \textbf{365}, 614 (1991).
\bibitem{Blaizot}
J.P. Blaizot, F.~Gelis, R.~Venugopalan, Nucl. Phys. A \textbf{743}, 13 (2004); F.~Gelis, Y.~Mehtar-Tani, Phys. Rev. D \textbf{73}, 034019 (2006),
\bibitem{Gelis:2010nm}
F.~Gelis {\it et al.}, 
Ann. Rev. Nucl. Part.
  Sci. \textbf{60}, 463 (2010), \texttt{1002.0333}
\bibitem{Raj:2023irr}
H.~Raj, R.~Venugopalan (2023), \texttt{2311.03463}
\bibitem{SabioVera}
  H.~Johansson {\it et al.}, 
  JHEP
  \textbf{10}, 215 (2013); A.~Sabio~Vera {\it et al.}, JHEP \textbf{04}, 086
  (2013).
\bibitem{Monteiro:2014cda}
R.~Monteiro, D.~O'Connell, C.D. White, JHEP \textbf{12}, 056 (2014).
\bibitem{Goldberger:2016iau}
W.D. Goldberger, A.K. Ridgway, Phys. Rev. D \textbf{95}, 125010 (2017),
\bibitem{Wong:1970fu}
S.K. Wong, Nuovo Cim. A \textbf{65}, 689 (1970)
\bibitem{Raj:2023iqn}
H.~Raj, R.~Venugopalan (2023), \texttt{2312.03507}
\bibitem{BFKL}
E.A. Kuraev, L.N. Lipatov, V.S. Fadin, Sov. Phys. JETP \textbf{45}, 199 (1977); 
I.I. Balitsky, L.N. Lipatov, Sov. J. Nucl. Phys. \textbf{28}, 822 (1978)
\bibitem{Mantysaari:2023gkw}
H.~M\"antysaari,  \texttt{2312.07805}
\bibitem{Dvali:2011aa}
G.~Dvali, C.~Gomez, Fortsch. Phys. \textbf{61}, 742 (2013).
\bibitem{Dvali:2019jjw}
G.~Dvali, Fortsch. Phys. \textbf{69}, 2000090 (2021).
\bibitem{Dvali:2021ooc}
G.~Dvali, R.~Venugopalan, Phys. Rev. D \textbf{105}, 056026 (2022).
  \bibitem{MV}
L.D. McLerran, R.~Venugopalan, Phys. Rev. D \textbf{49}, 3352 (1994); {\it ibid.}, Phys. Rev. \textbf{D49}, 2233 (1994); Phys. Rev. D \textbf{50}, 2225-2233 (1994).
  \bibitem{Coleman:1977hd}
S.~R.~Coleman,
Commun. Math. Phys. \textbf{55}, 113 (1977).
\bibitem{Duan:2021clk}
H.~Duan, A.~Kovner and V.~V.~Skokov, 
Phys. Rev. D \textbf{105}, no.5, 056009 (2022).
\bibitem{Bottomup}
R.~Baier {\it et al.},
Phys. Lett. B \textbf{502}, 51-58 (2001); J.~Berges {\it et al.}, Rev. Mod. Phys.
  \textbf{93}, 035003 (2021).
  \bibitem{String-Tension}
M.~Mace, S.~Schlichting and R.~Venugopalan,
Phys. Rev. D \textbf{93}, no.7, 074036 (2016); J.~Berges {\it et al.}, 
Phys. Rev. D \textbf{102}, no.3, 034014 (2020).
\bibitem{JIMWLK}
J.~Jalilian-Marian {\it et al.},
A.~Kovner, A.~Leonidov, H.~Weigert, 
  \textbf{504}, 415 (1997); E.~Iancu, A.~Leonidov, L.D. McLerran, Nucl. Phys. A \textbf{692}, 583 (2001).
 \bibitem{BK}
I.~Balitsky, Nucl. Phys. B \textbf{463}, 99 (1996); 
Y.V. Kovchegov, Phys. Rev. D \textbf{60}, 034008 (1999).
  \bibitem{Shock}
  A.~Ayala {\it et al.}, Phys. Rev. D
  \textbf{53}, 458 (1996); 
  I.I. Balitsky, A.V. Belitsky, Nucl. Phys. B \textbf{629}, 290 (2002). 
  \bibitem{Escriva:2022duf}
A.~Escriv\`a, F.~Kuhnel, Y.~Tada (2022), \texttt{2211.05767}
  
\end{thebibliography}
\end{document}